# Perfect absorption of a focused light beam by a single nanoparticle

Alexey Proskurin[1], Andrey Bogdanov[1], and Denis G. Baranov[2,3,*]

[1]*Department of Physics and Engineering, ITMO University, St. Petersburg 197101, Russia*

[2]*Department of Physics, Chalmers University of Technology, 412 96, Göteborg, Sweden*

[3]*Moscow Institute of Physics and Technology, Dolgoprudny 141700, Russia*

[*]Email: denisb@chalmers.se

**Abstract:** Absorption of electromagnetic energy by a dissipative material is one of the most fundamental electromagnetic processes that underlies a plethora of applied problems, including sensing and molecular detection, radar detection, wireless power transfer, and photovoltaics. Perfect absorption is a particular regime when all of the incoming electromagnetic energy is absorbed by the system without scattering. Commonly, the incident energy is delivered to the absorbing system by a plane wave, hence perfect absorption of this wave requires an infinitely extended planar structure. Here, we demonstrate theoretically that a confined incident beam carrying a finite amount of electromagnetic energy can be perfectly absorbed by a finite size deep subwavelength scatterer on a substrate. We analytically solve the self-consistent scattering problem in the dipole approximation and find a closed-form expression for the spatial spectrum of the incident field and the required complex polarizability of the particle. All analytical predictions are confirmed with full-wave simulations.

**Introduction**

The absorption of electromagnetic energy into a material is a phenomenon that underlies many applied problems, including molecular sensing, photovoltaics, and photodetection. The efficiency of absorption is the key parameter for those. Commonly, the incident energy is delivered to the system through a trivial single channel, such as a plane wave incident on one side of an absorber [1]. A classic example of an electromagnetic absorber is the Salisbury screen [2], consisting of a thin resistive sheet placed one quarter of the wavelength above a flat reflector. Perfect absorption can be realized in many other planar systems by the virtue of critical coupling, which requires equal radiative and dissipative decay rates of the system's eigenmode [3,4]

By exploiting the interference of multiple incident signals the electromagnetic absorption can be made more efficient and controllable/flexible [5–8]. A coherent perfect absorber (CPA) is a system in which complete absorption of electromagnetic radiation is achieved by the interference of several incident waves. The simplest CPA is a slab of an absorbing dielectric; when illuminated coherently from both sides by symmetric or anti-symmetric waveforms, it absorbs all incident energy. In this and related systems, the energy is delivered to the system with a plane wave, which, generally, requires the use of an absorber extended in two dimensions. It is possible to realize coherent perfect absorption in confined geometries, for example with transversely localized waveguide modes [9–11], or spheres and cylinders in a disordered environment [12] or free space [13]. In the latter scenario, irradiation with a CPA waveform leads to perfect absorption of the incident light by a localized surface plasmon. The main difficulty is that the incident CPA waves of a cylinder or a sphere are converging cylindrical or spherical waves, respectively, which contain a lot of evanescent components in their angular spectrum [14]. Similar ideas have been discussed previously in the context of perfect reflection of a focused beam by a dipolar particle [15]; partial extinction of a beam by a single molecule has been also observed [16]. The listed examples of perfect absorption mostly consider highly symmetry scatterers in homogeneous environment and neglect the substrate effect almost unavoidable in photonic problems. Such simplified consideration can be applied only for a very limited number of practical problems.

Here, we demonstrate theoretically that a dipole particle placed on a substrate can perfectly absorb a focused vectorial light beam, which does not have any near-field components in its spatial spectrum. Therefore, such a beam can be created in the far-field by conventional optical devices, such as phase plates and spatial light modulators, and focused onto the nanoparticle, where it will be absorbed. We establish an analytical solution of the scattering problem, suggest a realistic system supporting this effect, and verify it with numerical full-wave simulations.

**Results**

The basic intuition behind our idea is the following: a homogeneous plane wave carries an infinite energy flux $P = \int \frac{1}{2}\text{Re}(\mathbf{E} \times \mathbf{H}^*)d\mathbf{s}$. A nanosphere having only a finite absorption cross-section approximately bounded by $\lambda^2$ (or a finite absorption length for a cylindrical scatterer) cannot absorb all the energy of a plane wave. However, if the incident field is focused down to the diffraction limit of $\left(\frac{\lambda}{2}\right)^2$, its cross-section becomes comparable to the maximal absorption cross-section of the nanoparticle, thus suggesting that the focused incident field can, in principle, be perfectly absorbed. A similar problem has been studied recently in Ref. [17] for a cylinder, which showed that absorption cross-section can be controlled by tailoring the excitation field profile.

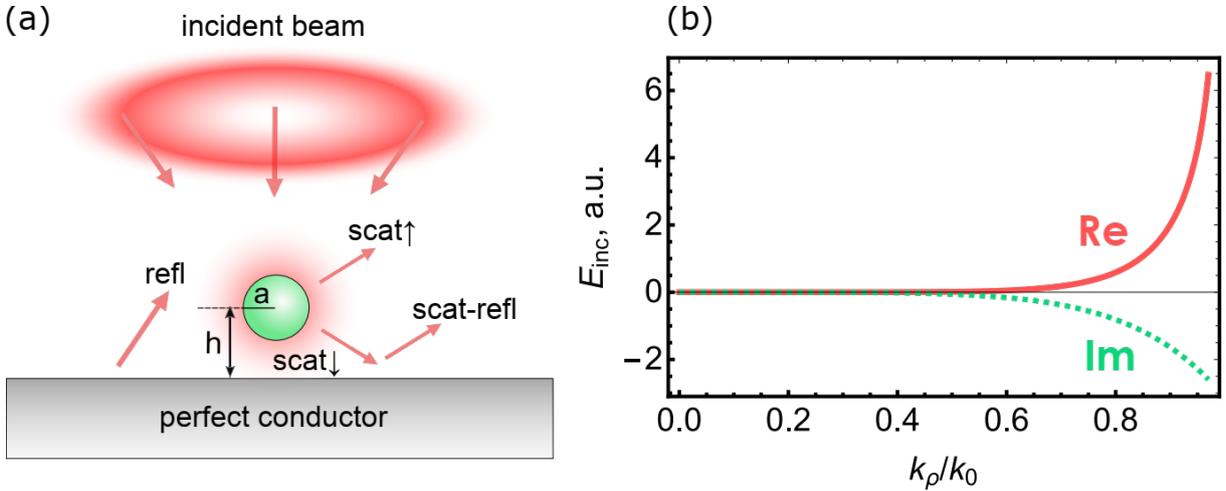

**Fig. 1.** (a) Schematic of the system under study representing various scattering pathways engaged in the problem. (b) Angular spectrum of the perfectly absorbing incident field (see Eq. 1) for $h = \lambda/4$.

The system under study is illustrated in Fig. 1a. We start by considering an absorbing subwavelength sphere placed at a distance $h$ above a perfectly conducting substrate. We will assume that the response of the nanosphere is limited to an electric dipole resonance, which is a good assumption for a metallic or dielectric subwavelength particle.

Thanks to the cylindrical symmetry of the problem, the solution can be sought for as a combination of radially and azimuthally polarized cylindrical beams [18]. The azimuthally polarized beam has zero electric field on its axis and thus it does not interact with an electric dipolar scatterer. Therefore, we consider only radially polarized incident beams. An incident monochromatic electric field oscillating at a frequency $\omega$ can be written in cylindrical coordinates as

$$\mathbf{E}_{inc}(\mathbf{r}) \equiv \begin{pmatrix} E_\rho \\ E_\phi \\ E_z \end{pmatrix} = \int_0^{k_0} \tilde{E}_0(k_\rho) \begin{pmatrix} -\frac{ik_z}{k} J_0'(k_\rho \rho) \\ 0 \\ J_0(k_\rho \rho) \end{pmatrix} e^{ik_z z} dk_\rho, \qquad (1)$$

where $k_0 = \omega/c$, $k_z = -\sqrt{k_0^2 - k_\rho^2}$ since the incident beam has components propagating only in the negative z direction towards the sphere, and $\tilde{E}_0(k_\rho)$ is the angular spectrum of $\mathbf{E}_{inc}(\mathbf{r})$. This field represents a linear combination of (non-diffracting) Bessel beams with various $k_\rho < k_0$, each being a combination of p-polarized plane waves with a fixed $k_z$ and all possible $(k_x, k_y)$ satisfying $k_x^2 + k_y^2 = k_\rho^2$. The integration limit from 0 to $k_0$ is imposed to make sure the incident field contains only far-field components.

Let us consider only the z-component of the field, thus, the problem becomes effectively scalar. To solve the scattering problem self-consistently we will use the conventional multiple-scattering method [19–21]. The total scattered field can be written as a sum of three components (Fig. 1b): $E_{scat,tot} = E_{ref} + E_{scat} + E_{scat \to ref}$, where $E_{ref}$ is the field of the incident beam reflected directly by the substrate without any interaction with the cylinder, $E_{scat}$ is the field scattered by the cylinder as if it was in a homogeneous dielectric environment, and $E_{scat \to ref}$ is the scattered field additionally reflected back by the substrate. The above expression for the total field can equally be written both in $\mathbf{r}$ and $k$-space. The reflected field is obtained by applying the reflection operator to the incident beam:

$$E_z^{ref} = \int_0^{k_0} r(k_\rho) \tilde{E}_0(k_\rho) J_0(k_\rho \rho) e^{i\delta} e^{ik_z z} dk_\rho, \qquad (2)$$

where $r(k_x)$ is the Fresnel reflection coefficient of the PEC substrate and $\delta(k_x) = 2k_z h$ is the double phase delay between the cylinder and the substrate. Here, $k_z$ is positive since it stands for a positive phase delay.

The incident field $\mathbf{E}_{inc}$ excites a vertical electric dipole $\mathbf{p} = (0,0,p_z)$, whose radiated filed can be written as (in SI units)

$$E_z^{scat} = \frac{1}{\varepsilon_0}\left(k_0^2 + \frac{\partial^2}{\partial z^2}\right)\frac{e^{ik_0 R}}{4\pi R}p_z. \qquad (3)$$

Sommerfield identity $\frac{e^{ik_0 R}}{R} = i\int_0^\infty \frac{k_\rho}{k_z}J_0(k_\rho\rho)dk_\rho$ [22] allows us to rewrite the scattered field in a compact way:

$$E_z^{scat} = \frac{i}{4\pi\varepsilon_0}\int_0^{k_0} p_z \frac{k_\rho^3}{k_z}J_0(k_\rho\rho)e^{ik_z|z|}dk_\rho, \qquad (4)$$

where $k_z = \sqrt{k_0^2 - k_\rho^2}$ for $k_\rho < k_0$ and $k_z = i\sqrt{k_\rho^2 - k_0^2}$ for $k_\rho > k_0$. The scattered-reflected field, therefore, can be obtained by applying the reflection operation to the fraction of the field radiated by the cylinder towards the substrate:

$$E_z^{scat\to ref} = \frac{i}{4\pi\varepsilon_0}\int_{-\infty}^{+\infty} p_z \frac{k_\rho^3}{k_z}r(k_\rho)e^{i\delta}J_0(k_\rho\rho)e^{ik_z z}dk_\rho. \qquad (5)$$

The vertical electric dipole moment $p_z$ is the response to the *total* electric field at the position of the sphere:

$$p_z = \varepsilon_0\alpha_0\left(E_z^{inc} + E_z^{ref} + E_z^{scat\to ref}\right)_{\mathbf{r}=0}, \qquad (6)$$

where $\alpha_0$ is the bare dipole polarizability of the sphere. Resolving Eq. (6) with respect to $p_z$ we find the dressed polarizability, which relates the dipole moment to the incident field at the particle's position via $p_z = \varepsilon_0\hat{\alpha}E_z^{inc}(0)$:

$$\hat{\alpha} = \frac{1}{\xi}\left(1 + \frac{\int_0^{k_0}\tilde{E}_0(k_\rho)re^{i\delta}dk_\rho}{E_z^{inc}(0)}\right), \qquad (7)$$

where $\xi = \frac{1}{\alpha_0} - \frac{i}{4\pi}\int_0^\infty \frac{k_\rho^3}{k_z}re^{i\delta}dk_\rho$ is a constant that does not depend on the incident field spectrum $\tilde{E}_0$.

Perfect absorption arises when all components of the total scattered field vanish for all propagating channels $k_\rho < k_0$: $\tilde{E}_{ref}(k_\rho) + \tilde{E}_{scat}(k_\rho) + \tilde{E}_{scat\to ref}(k_\rho) = 0$. Note that we do not require vanishing of scattered components with $k_\rho > k_0$: these evanescent components do not carry energy along z direction, although do carry energy along x in the general case. However, if the substrate does not support propagating guided modes, such as in the case of a perfect electric conductor (PEC) substrate, these spectral components of the scattered field do not contribute to the energy transfer in the horizontal plane. Thus, we are looking for a specific geometry of the system and incident field spectrum $\tilde{E}_0(k_\rho)$ yielding the perfect absorption condition. After substituting the sphere's dipole moment $p_z$, we obtain an integral Fredholm equation of the second kind:

$$\tilde{E}_0(k_\rho) + \frac{i}{4\pi\xi}\int_0^{k_0}\hat{K}(k_\rho,k_\rho')\tilde{E}_0(k_\rho')dk_\rho' = 0, \qquad (8)$$

where $\hat{K}(k_\rho, k'_\rho) = \frac{k_\rho^3}{k_z}\frac{1+re^{i\delta}}{re^{i\delta}}(1+r'e^{i\delta'})$, and variables with prime indicate that they are calculated as a function of $k'_\rho$. Thanks to the degenerate kernel, the problem admits an analytical solution. Indeed, according to Fredholm alternative [23], Eq. (8) has a nontrivial solution if and only if

$$\xi = -\frac{i}{4\pi}\int_0^{k_0}\frac{k_\rho^3}{k_z}\frac{(1+re^{i\delta})^2}{re^{i\delta}}dk_\rho. \tag{9}$$

The corresponding solution satisfying the integral equation with this $\xi$ therefore is (in arbitrary units)

$$\tilde{E}_0(k_\rho) \propto \frac{k_\rho^3}{k_z}\frac{(1+re^{i\delta})}{re^{i\delta}}, \tag{10}$$

and the bare polarizability of the sphere supporting perfect absorption for the given incident field is

$$\alpha_0 = 4\pi i\left(\int_0^{k_0}\frac{k_\rho^3}{k_z}\frac{(1+re^{i\delta})^2}{re^{i\delta}}dk_\rho - \int_0^\infty\frac{k_\rho^3}{k_z}re^{i\delta}dk_\rho\right)^{-1}. \tag{11}$$

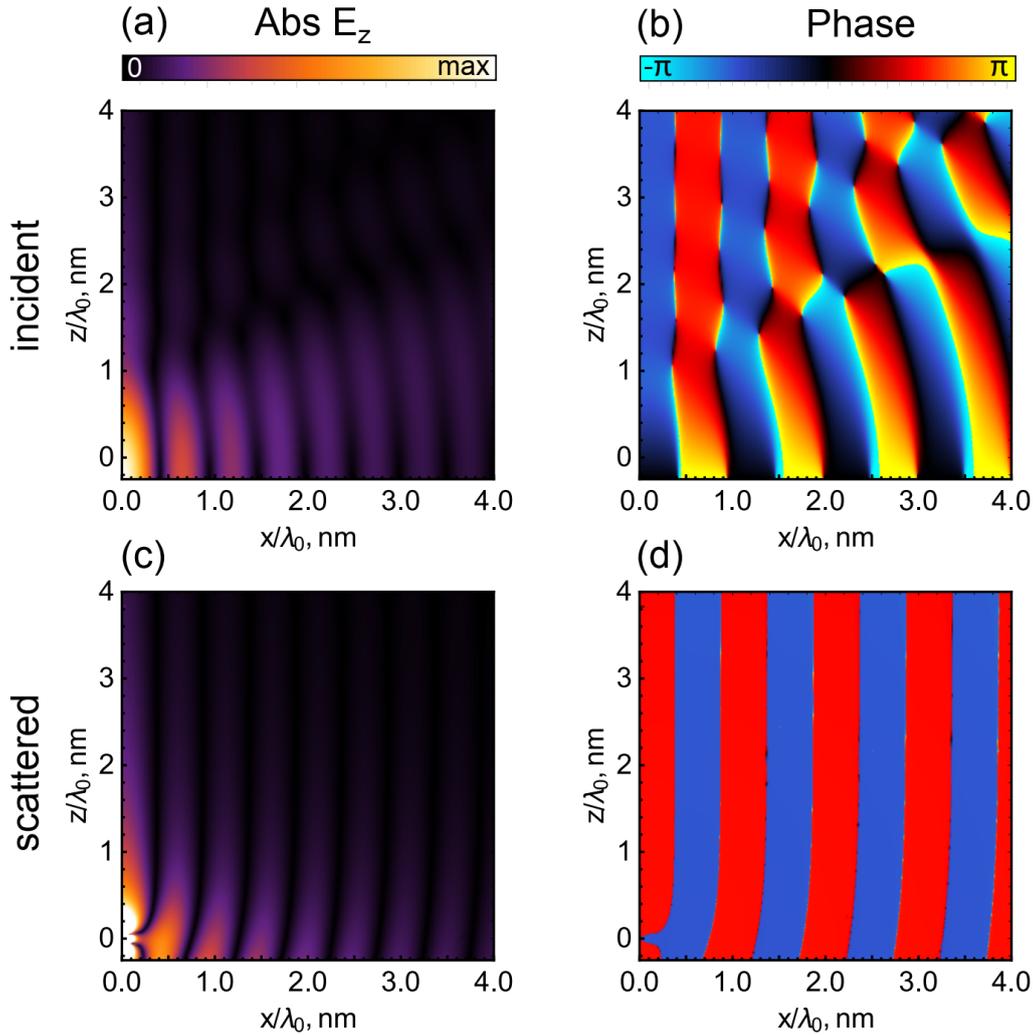

**Fig. 2.** Electric field distribution of a perfectly absorbing solution. (a,b) The absolute value and phase of the $E_z$ electric field component of the incident field satisfying the perfect absorption condition for $h = \lambda/4$ and a PEC substrate. The substrate is located at $z = -h$, the absorbing sphere is at $\mathbf{r} = 0$. (c,d) The same as (a,b) for the total scattered field.

Figure 1b shows the resulting angular spectrum $\tilde{E}_0(k_\rho)$ of the incident field given by Eq. (10) for a PEC substrate ($r(k_\rho) = 1$; bear in mind that $r$ relates z-components of the electric field) and $h = \lambda/4$. Interestingly, it shows that for a perfect absorption, most of the energy should be delivered to the sphere by harmonics with $k_\rho \sim k_0$ propagating at large oblique angles. This is in agreement with the fact that a vertical dipole radiates mostly in the horizontal plane. Corresponding spatial distributions of the incident and total scattered electric fields are plotted in Fig. 2. The argument of the incident field clearly reveals a phase gradient indicating the energy transfer from infinity towards the sphere. The phase of the scattered field, however, exhibits no gradient whatsoever, manifesting lack of energy transfer away from the scatterer.

Equations (10)-(11) dictate not only the angular spectrum of incident field, but also the sphere's polarizability $\alpha_0$ required for perfect absorption. Figure 3a shows the real and imaginary parts of $\alpha_0$ as a function of the distance $h$ between the sphere and PEC substrate. At moderate to large distances the real part of the polarizability oscillates near zero, while the imaginary part approaches a positive constant. Interestingly, the value of this constant is exactly $3\pi i \left(\frac{\lambda}{2\pi}\right)^3$ (as one can verify by integrating Eq. (11) in the limit $h = \infty$), i.e., the polarizability of a critically coupled dipolar scatterer having equal scattering and absorption free-space cross-sections [24,25]. This is a remarkable result, since this is exactly the condition for perfect absorption of a spherical harmonic by a resonant dipolar scatterer [25].

That and nearly zero (compared to $\lambda_0^3$) real part of the polarizability clearly indicates that a resonant scatterer is required to perfectly absorb the impinging beam. The required electric dipole polarizability $\alpha_0$ can be realized either with a subwavelength negative permittivity nanosphere, or with a Mie-resonant dielectric particle [26]. We choose a negative permittivity nanosphere as it is better approximated by a pure electric dipole scatterer. We can consider a few different options for the sphere's material: either a metal giving rise to plasmonic resonance, or a polar crystal giving rise to a phonon-polariton resonance [27]. To study the effect of the material's permittivity on the perfect absorption regime, we show in Fig. 3(b) the variation of $|\alpha_0 - \alpha_e|$ with real and imaginary parts of the sphere's permittivity, where $\alpha_e =$ $\left(\frac{1}{\alpha_{C-M}} - i\frac{k^3}{6\pi}\right)^{-1}$ is the corrected electric dipole polarizability with $\alpha_{C-M} = 4\pi a^3 \frac{\varepsilon-1}{\varepsilon+2}$ being the Clausius-Mossotti polarizability of a sphere of a radius $a$ [28]. The map has been calculated with a fixed $h = \lambda_0/4$ and $a/\lambda_0 = 0.02$. When $\alpha_e$ hits exactly the analytical value $\alpha_0$ at some $\text{Re } \varepsilon + i \text{ Im } \varepsilon$, the perfect absorption condition becomes satisfied, which can be seen as a dip in Fig. 3(b).

In order to evaluate if this regime can be reached with existing materials, we show parametric trajectories of complex permittivities yielding the perfect absorption condition parametrized with $a/\lambda_0$ in Fig. 3(c). One can see that these trajectories cross complex-valued permittivities of silver (adopted form [29]), aluminum (adopted from [30]), and SiC (adopted from [31]), chosen for demonstration here. To design the structure supporting the perfect beam absorption, one therefore should fix the wavelength at the point where the solution trajectory crosses the material's permittivity, and scale the sphere radius and the sphere-substrate distance accordingly.

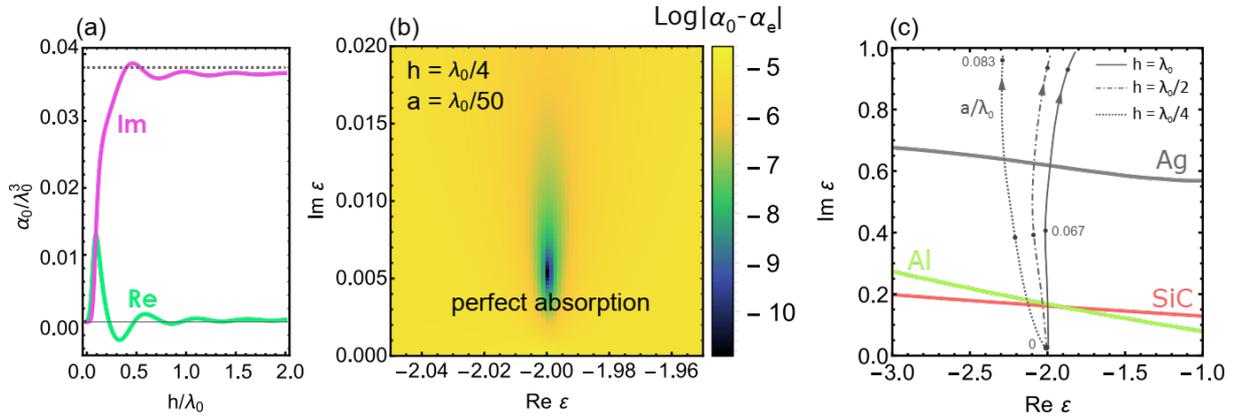

**Fig. 3.** (a) Bare electric dipole polarizability $\alpha_0$ required for perfect absorption (Eq. 11) as a function of the sphere-to-substrate distance $h$ for a PEC substrate; dashed line – polarizability of a critically coupled dipolar scatterer $3\pi i \left(\frac{\lambda_0}{2\pi}\right)^3$. (b) Logarithmic plot of $|\alpha_0 - \alpha_e|$ as a function of the sphere's permittivity for $h = \lambda_0/4$ and $a = 0.02\lambda_0$, where $\alpha_0$ is the polarizability of a perfectly absorbing sphere given by the analytical solution, and $\alpha_e$ is the is the corrected Clausius-Mossotti polarizability. (c) Trajectories of complex permittivities yielding the perfect absorption condition parametrized with $a/\lambda_0$ for a series of values of $h/\lambda_0$. Thick lines denote complex permittivities of silver, aluminum, and silicon carbide, crossing the analytical solution in specific points.

Next, we verified the analytical solution with full-wave finite-element method (FEM) simulations by using COMSOL Multiphysics® software [32]. We modeled scattering of an incident field given by Eq. (10) by a spherical particle of permittivity $\varepsilon$ and radius $a$ placed at a distance $h$ above a PEC substrate. Time-averaged absorption rate $W = \frac{\omega}{2}\int \text{Im}\,\varepsilon\,|\mathbf{E}(\mathbf{r})|^2 dV$ normalized by the total energy flux of the incident beam $I = \int \frac{1}{2}\text{Re}(\mathbf{E} \times \mathbf{H}^*)d\mathbf{s}$ (where the integration is carried over an infinite horizontal plane) calculated for a range of the particle's permittivity, Fig. 4(a), reveals a maximum as high as 0.94. Position of the observed absorption peak is close to the one predicted by the analytical dipole approximation for the same geometry (Fig. 3(b)). Figure 4(b) shows the spatial distribution of the total scattered field obtained at the point of maximal absorption. Discrepancies with the analytical results are most likely caused by the finite simulation area and higher multipole contributions.

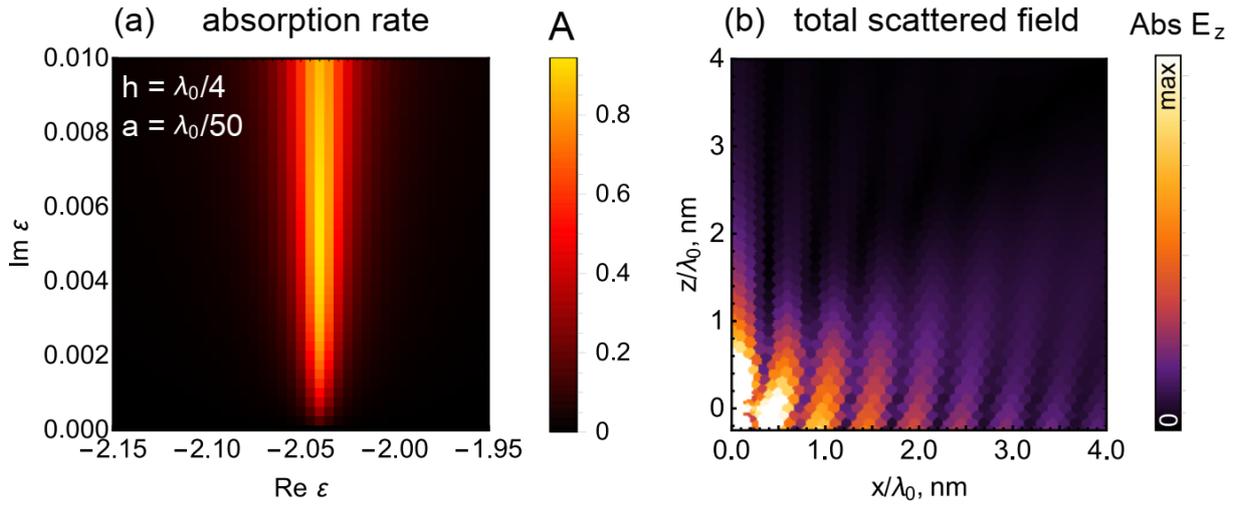

**Fig. 4.** Verification of the perfect absorption with full-wave FEM simulations. (a) Normalized absorption rate by a spherical particle on a PEC substrate illuminated by the perfectly absorbing beam, Eq. (10), as a function of the real and imaginary parts of the particle's permittivity for $h = \lambda_0/4$ and $a = 0.02\lambda_0$. (b) The z-component of the total scattered field for the sphere with the permittivity $\varepsilon = -2.04 + 0.005i$. Other parameters are the same as in panel (a).

We finally note that identical regime can be reached with a cylindrical scatterer (having an infinitely long axis) on a PEC substrate. Instead of illumination with radially polarized Bessel beams, the cylinder should be illuminated with a TE ort TM linearly polarized beam. For TM polarization and an electric dipole response, the problem becomes scalar and can be solved easily (see Supporting Information).

**Discussion and conclusion**

The demonstrated effect of perfect absorption bears a close connection to the problem of the dipole emission and time reversal symmetry. It is well known that the perfect absorption is a time-reversed process of lasing, which is the emission of coherent radiation [7]. In this context, time reversion has been used for subwavelength focusing in the far field [33,34]. In the problem addressed here, the obtained angular spectrum of the incident beam and the particle's polarizability represent a stationary monochromatic solution. Applying the time reversal operator yields another stationary solution. The stationarity of the time-reversal process is ensured by the proper polarizability of the particle precisely balancing the total incoming power carried by the fields and work performed by the field on the induced currents. Therefore, the incident beam required for perfect absorption can be found by reversing the far-field of a dipole placed above the metallic substrate, whereas the required polarizability of the dipole ensures the stationarity of the solution.

An oscillating dipole, however, produces both near and far fields. It might appear that in order to realize perfect absorption both the near- and far fields should be reversed. However, it is easy to see that *time reversal of the near field does not modify the field*. Indeed, the near field of a vertical dipole above a substrate is a cylindrically symmetric combination of evanescent waves with all possible $|\mathbf{k}_\parallel = (k_x, k_y)| > k_0$ and imaginary $k_z$. Reversing this field in time corresponds to complex conjugation, flipping the sign of $\mathbf{k}_\parallel$ of each spectral component, but maintaining the same $k_z$ [35]. Therefore, this operation yields the near field identical to the initial one. The same argument applies to the near field of a linear dipole considered in Supporting Information. In other words, converging (absorbing) and diverging (radiating) counterparts of the near-field part of a dipole emission are equivalent and do not require time-reversal.

To conclude, we have demonstrated that a focused incident beam containing only far-field propagating components in its spatial spectrum can be ideally absorbed by a localized point scatterer located above a reflective substrate. We have found an analytical solution of the scattering problem in the dipole approximation, which provides the spectrum of the incident beam and the required polarizability of the absorbing particle. We have also verified the effect with full-wave simulations. Our findings significantly expand the class of the perfect absorption phenomena and offer a new tool for electromagnetic energy harvesting. An interpretation of the effect in terms of time reversal operation also provides a simple way to generalize the perfect absorption to cases of arbitrary multipole excitations and substrates.

**Acknowledgements**

D.G.B. acknowledges support by the Russian Science Foundation grant 19-79-00362. A.A.B. acknowledge the RFBR (18-32-20205) and the Grant of the President of the Russian Federation (MK-2224.2020.2).